\documentclass{jpconf}
\usepackage{graphicx}
\usepackage{amsmath}
\usepackage{amssymb}
\usepackage{lmodern}

\begin{document}
\title{On the mixing time in the Wang-Landau algorithm}
\author{Marina Fadeeva$^{1, 2}$, Lev Shchur$^{1, 2, 3}$}
\address{$^1$ Science Center in Chernogolovka, 142432 Chernogolovka, Russia} 
\address{$^2$ National Research University Higher School of Economics, 101000 Moscow, Russia}
\address{$^3$ Landau Institute for Theoretical Physics, 142432 Chernogolovka, Russia}

\begin{abstract}
We present preliminary results of the investigation of the properties of the Markov random walk in the energy space generated by the Wang-Landau probability. We build transition matrix in the energy space (TMES) using the exact density of states for one-dimensional and two-dimensional Ising models.
The spectral gap of TMES is inversely proportional to the mixing time of the Markov chain. We estimate numerically the dependence of the mixing time on the lattice size, and extract the mixing exponent.
\end{abstract}

\section{Introduction} 

Wang-Landau algorithm is the way of the direct estimation of the density of states (DOS)~\cite{Wang-Landau,Wang-Landau-PRE}. 
It can be applied to any system with a defined partition function. The idea is based on the representation of the partition function 
as a sum over the energy levels 
\begin{equation}
Z=\sum_{k=1}^{N_E}g(E_k)e^{-E_k/k_BT}\;,
\label{partition-function}
\end{equation}
where $g(E_k)$ is the number of states (density of states) with the energy $E_k$, $N_E$ is the number of energy levels,
$k_B$ is the Boltzmann constant, and $T$ is the temperature.

It is proven to work quite well for many systems, with the examples in different areas of statistical physics and mechanics.
There are more than 15 hundreds of papers on the application of the algorithm and its improvements. 
In the paper we focus on the discussion of the mixing time in the final stage of the Wang-Landau algorithm. It is connected to the rate
of convergence of the estimated DOS. 

At the final stage of the correct Wang-Landau algorithm, the density of states is close to the true values~\cite{BFS-TM,Zhou2008}. Accordingly,
the random walk in the energy space with the Wang-Landau probability of acceptance of the new energy
\begin{equation}
P_{WL}(E_k,E_m)=\min\left(1,\frac{g(E_k)}{g(E_m)}\right)
\label{PWL-expr}
\end{equation}
becomes a reversible Markov process. A random walk in the energy space is characterized by the
transition matrix in the energy space (TMES)~\cite{BFS-TM}
\begin{equation}
T(E_k,E_m)=P_{WL}(E_k,E_m)P(E_k,E_m)
\label{TMES-expr}
\end{equation}
where $P(E_k,E_m)$ is a probability of one step of the random walk to move from a configuration with energy $E_k$
to any configuration with energy $E_m$. The TMES matrix is the stochastic symmetric matrix if computed with the true density of states $g(E_k)$.

We are interested in the mixing properties of the Markov random walk in the energy space generated by transitions from one energy level to a different one with the Wang-Landau probability. It is known~(see, f.e.~\cite{Boyd2004}) that mixing time of the Markov process is inversely proportional to the spectral gap $G$, which is the difference of the two largest eigenvalues
$\lambda_1>\lambda_2$ of the transition matrix
\begin{equation}
G=\lambda_1-\lambda_2 \;.
\label{gap}
\end{equation}

We present here a preliminary estimation of the spectral gap for one- and two-dimensional Ising models. The Ising model plays in statistical physics the same role as the fruit flies in biology. Ising model~\cite{Ising} is the simplest model of magnetics, and is the first model for which a problem of a continuous phase transition was solved in detail~\cite{Onsager}. 

In section~\ref{sec-1d} we present the estimation of the spectral gap and scaling of the mixing time for a one dimensional Ising model, for which we know exactly both the density of states and the elements of transition matrix for any length of the Ising spin chain. In section~\ref{sec-2d} we calculate the transition matrix numerically for the 2D Ising model for which we know exactly the density of states~\cite{Beale}.  We estimate the gap dependence on the system size, and present a numerical estimate of the mixing time scaling with the system size. We compare our results with the known estimates of the tunnelling time in a 2D Ising model. In the final section~\ref{sec-discussion} we discuss future work.
 
\section{Mixing time for one-dimensional Ising model}
\label{sec-1d}

Consider a one-dimensional Ising model with $L$ spins $\sigma_i$ ($i=1,2,...,L$) placed on the string with periodic boundary conditions. The energy is given by 
\begin{equation}
E=-\sum_{i=1}^{L}J\sigma_i \sigma_{i+1}
\label{energy}
\end{equation}
and $\sigma_{L+1}=\sigma_1$. Without loss of generality and for the sake of simplicity we can put $J=1$ (in other words, we measure energy $E$ in units of $J$). Energy~(\ref{energy}) can be parameterized by $k$, the number of pairs of the domain walls, $E_k=-L+4k$. Occupations of energy levels are nothing but binomial coefficients, therefore the density of states is $g(E_k)=2C_L^{2k}$. Therefore, off-diagonal matrix elements of TMES are~\cite{BFS-TM}
\begin{equation}
T(E_k,E_{k+1})=T(E_{k+1},E_k)=\frac{C^{2k}_{L-2}}{\max\left(C^{2k}_L,C^{2k+2}_L\right)}.
\label{tmatrix1d}
\end{equation}
More details can be found in the paper~\cite{BFS-TM}.

\begin{figure*}
\center
\includegraphics[width=.5\textwidth, keepaspectratio=True]{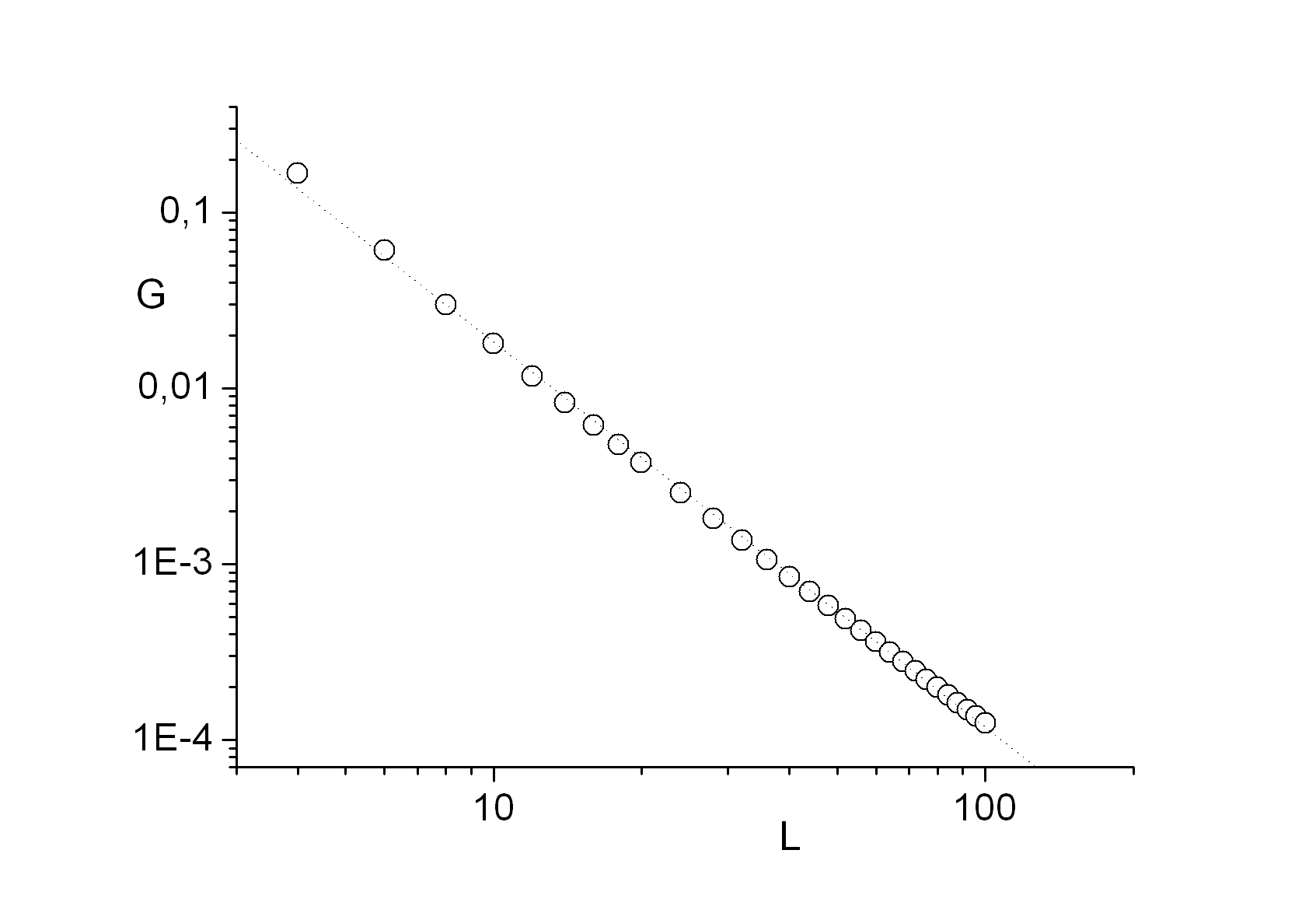}~\includegraphics[width=.5\textwidth, keepaspectratio=True]{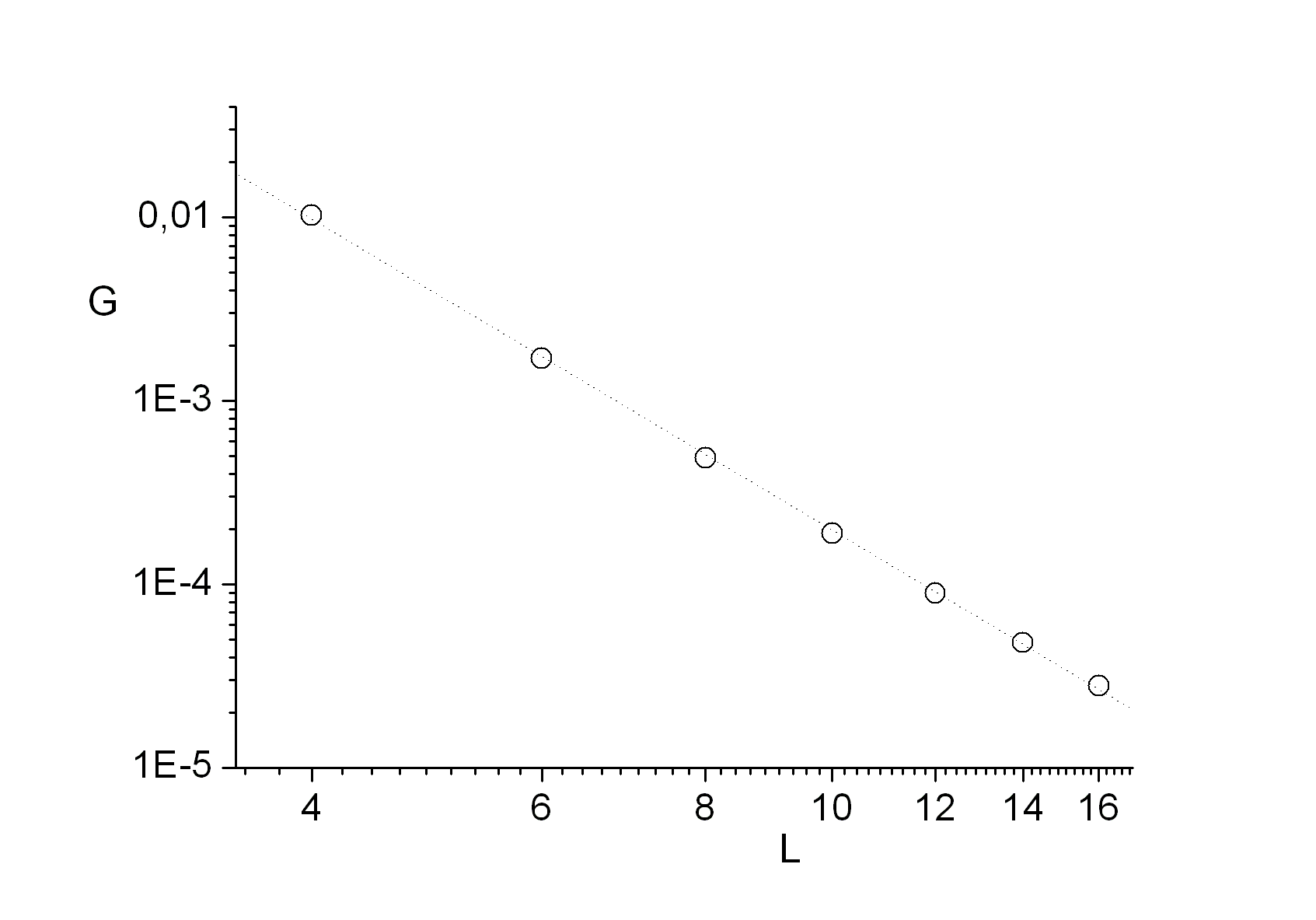}
\caption{Spectral gap $G$ (circles) as a function of the linear lattice size $L$. Dotted line is the linear fit to $G$. Left: one-dimensional Ising model. Right: two-dimensional Ising model. Note the log-log scale.
}
\label{fig:G-1d}
\end{figure*}

Using this expression, we compute elements of TMES for the number of spins $L$ ranging from 4 to 100, and compute two largest eigenvalues using the $\tt dgev$ function from the LAPACK~\cite{LAPACK} library. Dependence of the gap on $L$ is shown in the left panel of Figure~\ref{fig:G-1d} in the logarithmic scale for both axes, which clearly demonstrates the power law decay of the spectral gap. The linear fit gives the slope value $-2.19(2)$. Therefore, the mixing time $T_m=1/G$ in the Wang-Landau algorithm for one-dimensional Ising model scales with the system size $L$ as $T_m \propto L^{2.19}$.

\section{Mixing time for two-dimensional Ising model}
\label{sec-2d}

Matrix elements of TMES for the two-dimensional Ising model can be computed using the Wang-Landau probabilities~(\ref{PWL-expr}) with the exact values of DoS~\cite{Beale}. We compute TMES for the system sizes $L$ from 4 to 16. We have to stress that convergence of the TMES spectrum is very slow, and is slowing down with the system size $L$. For example, the eigenvalue $\lambda_2$ reaches its stationary value after approximately $10^6$ WL-steps for $L=4$, after $10^{8}$ steps for $L=8$, and after $10^{12}$ steps for $L=16$. The linear fit to the $G(L)$ data in the log-log scale gives slope of the line $-4.28(4)$. Therefore, $T_m\propto L^{4.28}$ for the two-dimensional Ising model.

In the paper~\cite{Dayal2004}, the average {\em tunneling time} $\tau$ to move from a configuration state with the lowest energy to the configuration with the highest energy has been measured for the two-dimensional Ising model, and was found to scale as $\tau\propto L^{4.743(7)}$.  The number $N_E$ of energy levels for this model is $N_E\propto L^2$.  The tunnelling time for the unbiased random walk in the energy space will be $\tau_0\propto N^2$. The bias due to the WL probability for the random walk does increase this time. 

\section{Discussion}
\label{sec-discussion}

In the paper we estimate the mixing time of the random walk in the energy spectrum generated by the Wang-Landau probability. Using the approach proposed in the paper~\cite{BFS-TM}, we compute the transition matrix in the energy spectrum (TMES) for one-dimensional and two-dimensional Ising models using the exact knowledge of the density of states. We estimate the spectral gap of TMES, and obtain the exponents of the mixing time $T_m$. This time can be viewed as the characteristic time at the final stage of a $1/t$ Wang Landau algorithm~\cite{BP-1,BP-2} (or a stochastic approximation algorithm~\cite{SAMC-1,SAMC-2}).

The connection of the mixing time we measure in the paper and the tunnelling time~\cite{Dayal2004} for the Markov process is not obvious. Future work is necessary for the analysis of performance limitations of the Wang-Landau algorithm.

\ack
This work was supported by grant 14-21-00158 from the Russian Science Foundation.


\end{document}